# Spin-wave-induced lateral temperature gradient in a YIG thin film/GGG system excited in an ESR cavity

Ei Shigematsu, Yuichiro Ando, Sergey Dushenko, Teruya Shinjo, and Masashi Shiraishi

*Department of Electronic Science and Engineering, Kyoto University, 615-8510, Kyoto, Japan*

**Lateral thermal gradient of an yttrium iron garnet (YIG) film under the microwave application in the cavity of the electron spin resonance system (ESR) was measured at room temperature by fabricating a Cu/Sb thermocouple onto it. To date, thermal transport in YIG films caused by the Damon-Eshbach mode (DEM)—the unidirectional spin-wave heat conveyer effect—was demonstrated only by the excitation using coplanar waveguides. Here we show that effect exists even under YIG excitation using the ESR cavity—tool often employed to realize spin pumping. The temperature difference observed around the ferromagnetic resonance (FMR) field under the 4 mW microwave power peaked at 13 mK. The observed thermoelectric signal indicates the imbalance of the population between the DEMs that propagate near the top and bottom surfaces of the YIG film. We attribute the DEM population imbalance to the different magnetic damping near the top and bottom YIG surfaces. Additionally, the spin wave dynamics of the system were investigated using the micromagnetic simulations. The micromagnetic simulations confirmed the existence of the DEM imbalance in the system with the increased Gilbert damping at one of the YIG interfaces. The reported results are indispensable for the quantitative estimation of the electromotive force in the spin-charge conversion experiments using ESR cavities.**

Spin caloritronics[1]—young but quickly developing spintronics field—is in pursuit of the comprehensive understanding of the connection between heat and spin currents. Following the discovery of the spin Seebeck effect[2], plenty of experimental demonstrations of spin caloritronic phenomena have been reported, such as the spin-dependent Seebeck effect[3] and the spin Peltier effect[4]. Especially, the heat transport via spin waves and spin-phonon interaction has attracted attention after the unidirectional spin-wave heat conveyer effect was reported[5,6]. In contrast to the conventional case of the heat transport against the temperature gradient, Damon-Eshbach mode (DEM) spin wave[7,8] induces heat transport in the direction of the thermal gradient. Apart from this surprising achievement, the unidirectional spin-wave heat conveyer effect has important implications in the field of the dynamical spin injection, also known as spin pumping, since they often occur simultaneously in a studied system. Spin pumping[9,10] is a method of generation of the pure spin current in the material due to the coupling of the interface spins to the precession magnetization of the adjacent ferromagnet layer. It quickly gained popularity as a spin injection method that can easily be used in any bilayer system consisting of nonmagnetic/ferromagnetic material[11,12] whereas the electrical spin injection needs more elaborate surface treatment, formation of tunnel barriers and nanofabrication[13–15]. Using the spin pumping technique, the spin-to-charge conversion-related properties of various heavy metals[11,16], semimetals[17–19], semiconductors[20–22] and even two-dimensional materials[23–25] were unveiled, along with the spin transport properties of the materials [24,26,27].

The DEM is the surface spin wave that is excited under the conditions close to the ferromagnetic resonance (FMR) and propagates in the opposite directions on the top and bottom surfaces[28,29]. When DEM reaches the end of the sample, its energy is damped as the heat, raising the temperature near the sample edge. In the case of the uniform excitation across the ferromagnet, the population of the DEM on top and bottom surfaces is the same, and the net quantity of the transported heat cancels out. However, when the equivalence of the population of the two DEM propagating in the opposite directions is broken, the unidirectional thermal transport takes place. In the previous studies of the unidirectional spin-wave heat conveyer effect[5,30], such inequivalence was shown to be present in case of the DEM excitation using the microstrip lines waveguides. In that case, the bottom surface of the ferromagnet is located closer to the microstrip line than the top surface, thus difference in the intensity of the microwave AC magnetic field causes population difference of the two DEM spin waves. Induced unidirectional heat transport happens in the direction of the propagation of the dominant DEM. Importantly, direction of the propagation of the dominant DEM (wave vector **k**) can be reversed by reversing the direction of the external magnetic field. Thus, voltage generated due to thermal effects (for example, the Seebeck effect) also reversed with the direction of the magnetic field. Incidentally, the spin pumping and spin-charge conversion experiments rely heavily on the reversal of the magnetic field to exclude non-magnetic spurious effects, including the thermal ones: sign reversal of the generated electromotive force with the external magnetic field usually taken as a proof of its spin-charge conversion origin. Thus, to confirm the origin of the electromotive force in the spin-charge conversion experiments, it is crucial to precisely estimate the unidirectional heat transfer induced by the DEM.

While there were a few experimental[5,30] and theoretical[31] studies on the unidirectional heat transfer effect under the microwave excitation using wave guides, there were no such reports in the microwave cavities. In contrast, broad variety of the spin pumping and spin-charge conversion experiments are carried out using the $TE_{011}$ cavity of the electron spin resonance (ESR) systems[11,32,33]. Our study fills the experimental gap, and reports the observation of the heat transfer by the DEM in the $TE_{011}$ ESR cavity. We also performed the micromagnetic simulations, and discuss the origin of the DEM imbalance observed experimentally.

We now proceed to the experimental details and results. The 1.2-μm-thick YIG film was grown by liquid phase epitaxy on top of the GGG substrate and is available commercially (Granopt, Japan). We fabricated thermocouple on top of the YIG surface to measure temperature difference generated due to the heat transport by the DEM. While there are many types of thermocouples available commercially, the most common ones (types E, J, K, T) use ferromagnetic metals nickel (Ni) and iron (Fe), or their alloys, which may exhibit ferromagnetism due to the insufficient uniformity of the alloy. In the spin pumping experiments the lateral static magnetic field is applied in plane of the samples under the ferromagnetic resonance condition. In this geometry, the anomalous Hall effect in the thermocouple may be induced by the heating of the YIG film, which would add up to the electromotive force generated by the lateral thermal gradient of YIG film and prevent its quantitative estimation. To realize a thermocouple comprised of nonmagnetic metals, we focus on the combination of copper (Cu) and antimony (Sb), and use Cu wiring to make an electrical contact to the sample. First, we formed a 50-nm-thick $SiO_2$ insulating layer on top of YIG to exclude the influence of spin pumping, which was shown to decrease exponentially with the thickness of the tunnel barrier[34] . On top of it, 50-nm-thick Sb layer was deposited by resistance heating deposition. Finally, the third layer consisting of two Cu pads separated by a 1 mm gap was deposited. The sample with the formed thermocouple was set in the Seebeck effect measurement system (Fig.1(a)). Room temperature acted as a baseline level, while the hot and cold heat sinks—the temperature of which was controlled by the Peltier elements—were attached to the opposite sides of the sample. Lateral temperature difference and thermoelectric electromotive force were monitored simultaneously. For the ferromagnetic resonance measurements, the sample was mounted into the $TE_{011}$ cavity of the ESR system (JEOL JES-FA200) at room temperature. The DC and AC magnetic fields were applied in plane of the sample in DEM geometry as shown in Fig.1(b). The frequency of the AC magnetic field was set to 9.12 GHz, and applied microwave power was set to 4 mW. An estimated value of AC magnetic field applied to the sample was 2.2 μT. The DC magnetic field was swept through the FMR field of the YIG film, while the microwave absorption spectrum and the electromotive force between Cu electrodes were measured simultaneously. Since we used a bipolar electromagnet, measurements in 0° and 180° DC magnetic field are carried out without rotating the sample position

Figure 1(a) and Figure 2 show the schematic layout and the detected thermoelectric electromotive force in the Seebeck effect measurement of the Cu/Sb thermocouple fabricated on top of the the YIG/GGG sample. We follow the conventional definition of the Seebeck coefficient $S$:

$$\Delta V = -S\Delta T. \qquad (1)$$

where $\Delta V$ and $\Delta T$ are the thermoelectric electromotive force and the temperature difference, respectively. From the linear fitting (black solid line in Fig. 2), the Seebeck coefficient of the fabricated sample was determined to be +15 nV/mK. This result is comparable to the Seebeck coefficient of amorphous Sb film reported in the literature[35].

Next, the sample was placed in the cavity of the ESR system for the measurement of the magnetic-field-dependent heat transport induced by the DEM. The ferromagnetic resonance measurements with simultaneous detection of the electromotive force and FMR spectra were carried for the opposite directions of the DC magnetic field 0° and 180°. The wave vector **k** of the DEM is parallel to the cross product of the DC component of the magnetization of the YIG film **M** and the normal vector to the surface **n**. Direction of the **k** determines the direction $\zeta_{\Delta T}$ of the generated temperature difference $\Delta T$ on the propagation surface[5,7],

$$\mathbf{k} \,//\, \zeta_{\Delta T} \,//\, \mathbf{M} \times \mathbf{n} \qquad (2)$$

Therefore, we extracted the magnetization-dependent component of the observed thermoelectric signal by subtracting the signals measured at 0° and 180° direction of external magnetic field ($V_{0°}$ and $V_{180°}$, correspondingly). Figure 3 shows the thermoelectric signal generated by the DEM, which is given by $V_m = (V_{0°} - V_{180°})/2$, and the FMR spectra at the DC magnetic fields of 0° and 180°. The coincidence of the two FMR spectra confirms the identical resonance conditions for the opposite directions of the DC magnetic field. Interestingly, the DEM thermoelectric signal shows reversal of the polarity when approaching the FMR condition. Following the results of the Seebeck effect measurement of the sample, positive $V_m$ signal corresponds to $+y$ direction of the thermal gradient $\zeta_{\Delta T}$, which is due to the DEM at the GGG/YIG interface, and negative $V_m$ to $-y$ direction, which is due to the DEM at the $SiO_2$/YIG interface, respectively. The amplitude of the negative peak of the thermoelectric signal was measured to be -190 nV. Using the Seebeck coefficient of the sample, the estimated temperature difference between the Cu pads separated by the 1 mm gap ($-y$ direction) is 13 mK. Figure 4 shows the schematic layout of the DEM excitation in our measuring geometry for 0° direction of the external magnetic field. The thermal gradient direction $\zeta_{\Delta T}$ of $-y$ ($+y$) suggests the contribution of the DEM from the YIG interface with the $SiO_2$ (the GGG) film. Note that the uniformly excited DEMs in the thin ferromagnetic film has the same population of the +**k** and -**k** modes, thus they transfer the equal amount of heat in the opposite directions and the induced temperature differences by the two modes cancel each other out. Therefore, the negative peak of the thermoelectric electromotive force at the DC magnetic field close to the FMR condition signifies that the magnitude of the DEM at the $SiO_2$/YIG interface is superior to that on the GGG/YIG interface.

The previous analytical magnetostatic studies of the DEM assumed that the ferromagnetic film was placed in the vacuum and did not treat the symmetry breaking of the top and bottom sides of the film[7]. Furthermore, the influence of the Gilbert damping on the DEM propagation and damping was not considered. We carried out numerical micromagnetic simulations that evaluate effect of the symmetry breaking in our $SiO_2$/YIG/GGG system on the DEM population using program MuMax[3] [36]. GGG is known as a paramagnetic material with substantially large magnetization. Influence of the GGG layer attached to the YIG interface on the Gilbert damping of the surface YIG layer was already pointed out[30]. Thus, in the micromagnetic simulations, we set the Gilbert damping parameter $\alpha$ of one marginal layer next to the YIG/GGG interface (we refer to it as the bottom layer) larger than the other layers. The Gilbert damping parameter of the bottom layer was 0.02 and that of the other layers was 0.002 (Fig. 5(a)). As for the other magnetic parameters, we use those of permalloy presented in the specification paper of MuMax[3] [36], as a simple magnetic thin film model. The saturation magnetization and the exchange stiffness were set to be $8.6\times10^{5}$ A/m and $1.3\times10^{-11}$ J/m, respectively. At the beginning of the simulation, the DC magnetic field was set, and the magnetization of the whole system was relaxed. Following that, the AC excitation of the magnetic field was applied, and—after the magnetization precession reached the steady state—we extracted the $z$ component of the magnetization of each spin cell in the slice of $x$ = 25 (where coordinate represents layer number in that direction). The magnetization motion in the slice consists of a non-time-dependent bias, standing waves, and traveling waves along the $y$ direction. We can evaluate the DEM by extracting a portion of the traveling waves. For this purpose, the Fourier transform was implemented:

$$m_z\left(f,\frac{k_y}{2\pi}\right)=\int_{y_{\min}}^{y_{\max}}\int_{t_{\min}}^{t_{\max}}\mathrm{W}(t)\cdot\mathrm{W}(y)\cdot m_z(t,y)e^{-2\pi\mathrm{j}\left(ft-\frac{k_y}{2\pi}y\right)}\mathrm{d}t\mathrm{d}y \quad (3)$$

where $m_z$, $W$, $f$, and $k_y$ denote the $z$ component of the normalized magnetization, the hamming window function, the excitation frequency, and the wavenumber of the magnetization in $y$ direction, respectively. As we extracted the data in the finite range of $[[y_{\min}, y_{\max}],[t_{\min}, t_{\max}]]$, we applied the hamming function to reduce obstructive sub lobes in the resulting Fourier spectra. We focus on the Fourier spectra in the frequency of the excitation $f_0$, which was set to 15 GHz. When $m_z\left(f_0, \frac{k_y}{2\pi}\right)$ is deconvoluted into $m_z\left(f_0, \frac{k_y}{2\pi}\right) = A + B\mathrm{j}$ and $m_z\left(-f_0, \frac{-k_y}{2\pi}\right) = C + D\mathrm{j}$, the absolute amplitude with wavenumber $k$ leads to $m_z^{\mathrm{abs}}\left(\frac{k_y}{2\pi}\right) = \sqrt{(A+C)^2 + (-B+D)^2}$. Then we obtain the wave distribution as a function of the wavenumber $k_y/2\pi$ . A plot of the amplitude $m_z^{\mathrm{abs}}$ vs. wavenumber at different DC magnetic fields is shown in Fig. 5(b). The extracted plot at the magnetic field of 325.6 mT is also shown in Fig. 5(c). The clear break of the symmetry can be seen between the wavenumber spectra in the top (red line) and bottom layers (green line). The peak height in the top layer was superior to that in the bottom layer. Figure 5(d) shows the absolute amplitude $m_z^{\mathrm{abs}}$ in the central layer (top box), which is analogous to the FMR spectrum detected experimentally; the subtraction of the $m_z^{\mathrm{abs}}$ between the top and bottom layers (middle box), which characterizes the imbalance of the DEM between them; and the wave number (bottom box) corresponding to the maximum $m_z^{\mathrm{abs}}$ in the top (red filled circles) and bottom (green filled circles) layers. The peak of the DEM in simulated spectra appeared at the lower magnetic field than the FMR, in agreement with theoretical and experimental results in the literature[37]. The DEM modes in the top and bottom layers had opposite sign of the wave number (Fig. 4(d) bottom), i.e. the propagation direction of the DEM, and were consistent with DEM propagation direction in the previous studies[5,30]. Thus, the numerical simulation showed the imbalance between the DEM in the top and bottom layers due to the difference in damping constant. Finally, we performed MuMax$^3$ simulations using parameters close to the experimental values. The AC magnetic field excitation frequency was set to $f_0$ = 9.12 GHz. The saturation magnetization of the YIG layer was set to be $1.275\times10^5$ A/m, and the exchange stiffness to $3.7\times10^{-12}$ J/m[38]. The calculated geometry is illustrated in Fig. 6(a). The Gilbert damping was 0.01 and 0.001 in the bottom layer and bulk, respectively. As in the case of permalloy, the DEM amplitudes in the top and bottom layers showed a clear difference (Fig. 6(b)). The difference in the amplitude between the two DEM propagating in the opposite directions peaked around the FMR resonance field (Fig. 6(c)). Thus, numerical simulation of YIG also indicated that the top-layer DEM is dominant over the bottom-layer DEM due to the break of the reflection symmetry because of the different magnetic damping at the surfaces. This result explains the dominance of the heat generated by the top DEM observed in the experiment. We note that micromagnetic calculations for the full-size sample are necessary for the precise quantitative simulation of the experimental results, which was limited by the computational power in this study. Additionally, the quantitative determination of the heat drift velocity—a key parameter in the thermal distribution induced by the DEM imbalance—needs a more elaborate analysis of the spin-phonon interaction, which is left for the further study. However, our experimental and numerical results clearly show that reflection symmetry breaking between the two YIG surfaces by the magnetic damping at the interfaces induces the imbalanced DEM population and the unidirectional heat transfer.

In conclusion, we observed the unidirectional spin-wave heat conveyer effect in a 1.2-μm-thick YIG film under the uniform microwave excitation in the ESR cavity. The origin of the DEM imbalance that led to the heat transport is explained by the increased Gilbert damping at one of the YIG interfaces. The micromagnetic simulations confirmed the existence of the DEM imbalance in such system. Our study fills the experimental gap that existed in the literature on the unidirectional spin-wave heat conveyer effect generated in ESR cavity. The reported results are indispensable for the quantitative estimation of the electromotive force in the spin-charge conversion experiments using ESR cavities.

**Supplementary Material**

The supplementary material includes the following information, microwave power dependence of the detected electromotive force, dependence of the detected electromotive force on the direction and speed of the DC magnetic field sweep, reproducibility of the results, and details of the MuMax3 calculation.

**Acknowledgements**

E.S. acknowledges the financial support from the JSPS Research Fellowship for Young Researchers and JSPS KAKENHI Grant No. 17J09520. This work was supported in part by MEXT (Innovative Area “Nano Spin Conversion Science” KAKENHI No. 26103003), Grant-in-Aid for Scientific Research (S) No. 16H06330, and Grant-in-Aid for Young Scientists (A) No. 16H06089. S.D. acknowledges support by JSPS Postdoctoral Fellowship and JSPS KAKENHI Grant No. 16F16064. The authors thank T. Takenobu and K. Kanahashi for the informative advice regarding the Seebeck coefficient measurement.

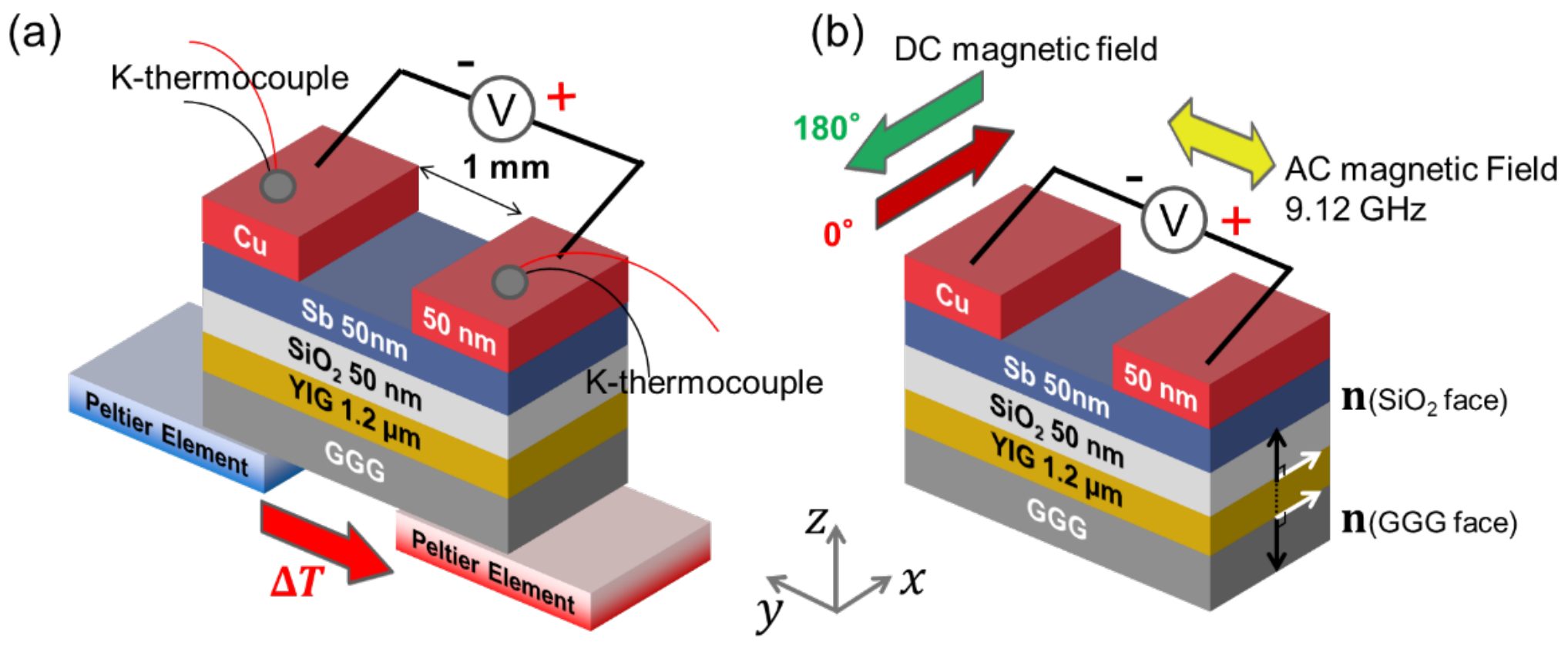

**Fig. 1.** (a) A schematic image of the Seebeck effect measurement. The fabricated sample was attached to the two Peltier elements that controlled the temperature difference between the edges of the sample. (b) A schematic image of the measurement of the DEM heat transfer under the FMR excitation in the ESR $TE_{011}$ cavity.

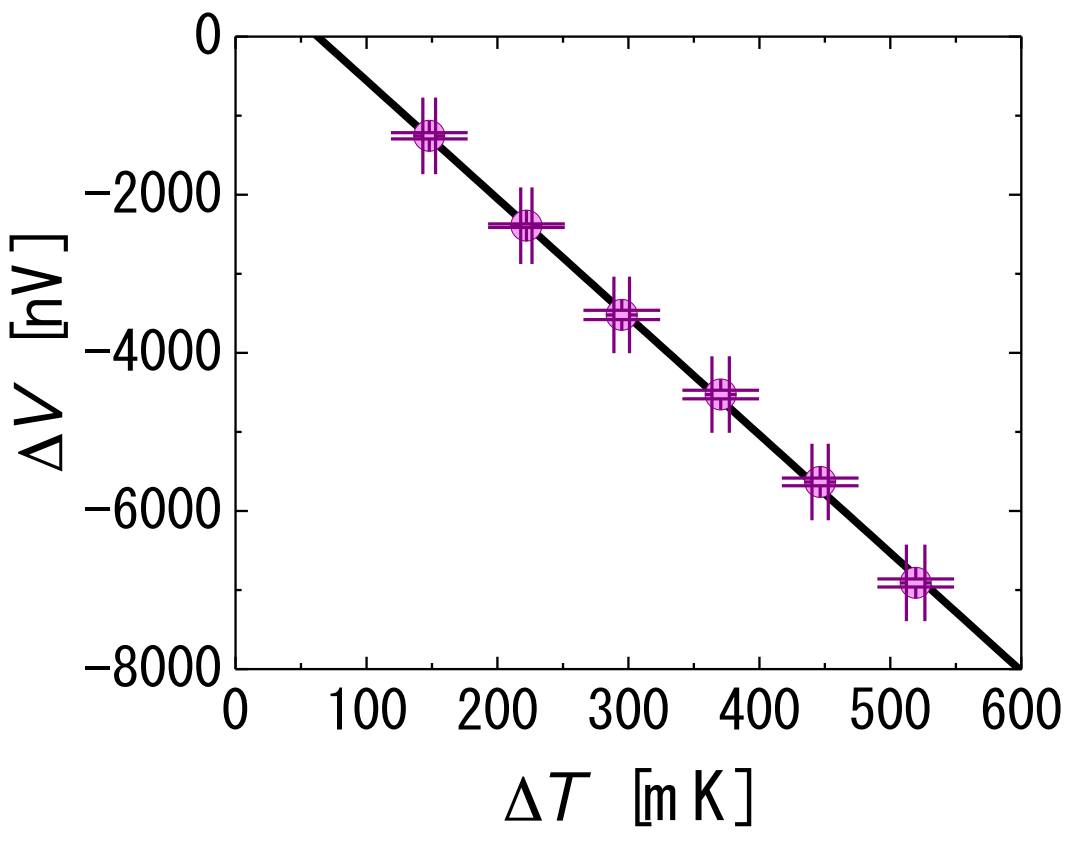

**Fig. 2**. The thermoelectric electromotive force dependence on the applied temperature gradient for the Sb/Cu thermocouple fabricated on top of the YIG film. The black solid line is a linear fitting.

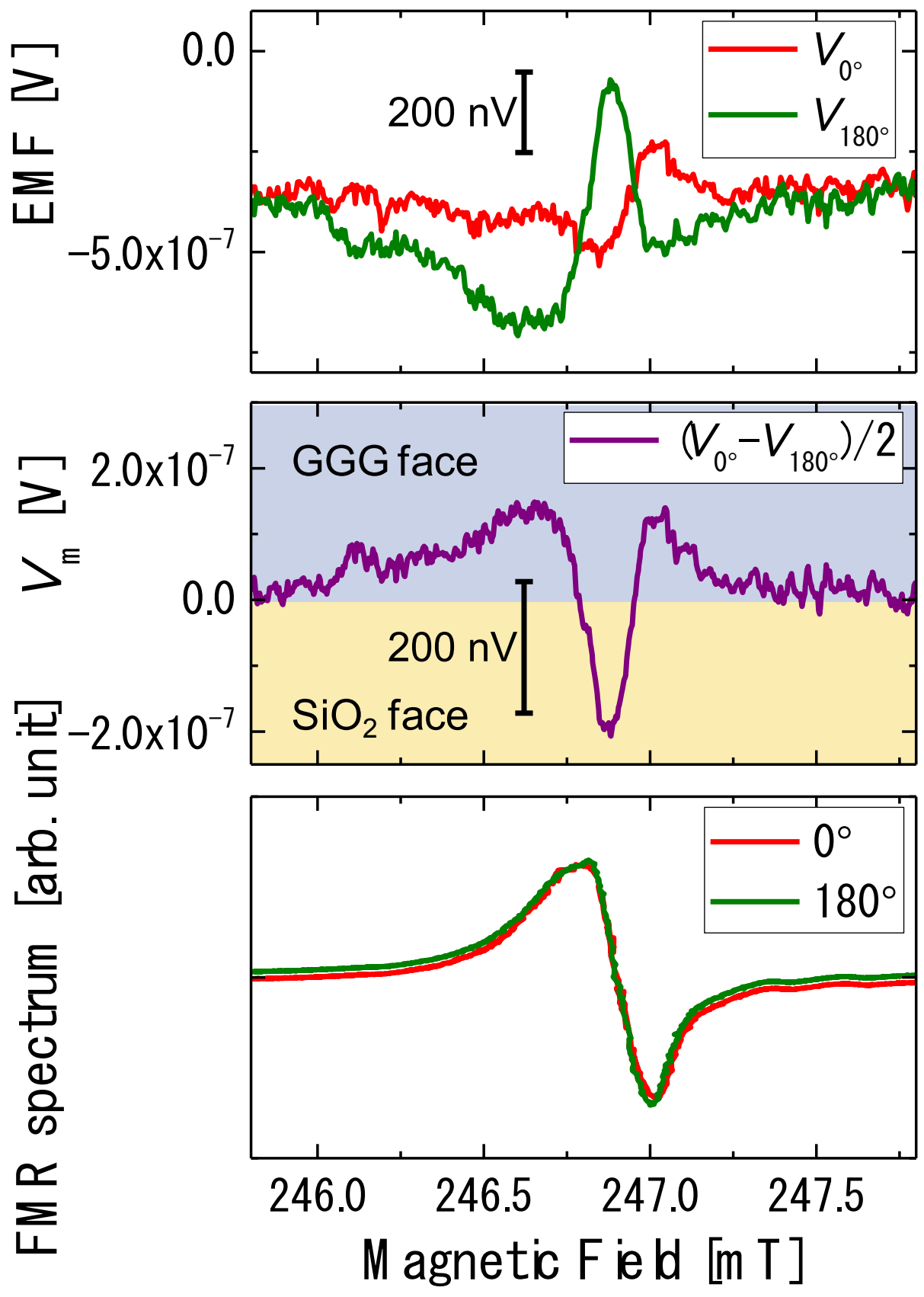

**Fig. 3**. Two lines are the electromotive forces observed under the microwave excitation of 4 mW at the DC magnetic fields of 0° and 180° A line in the middle box is the halved subtraction of the electromotive force ($V_{\mathrm{m}}$). Two overlapped lines in the lower box are FMR spectra measured for 0° (red) and 180° (green) direction of the DC magnetic field.

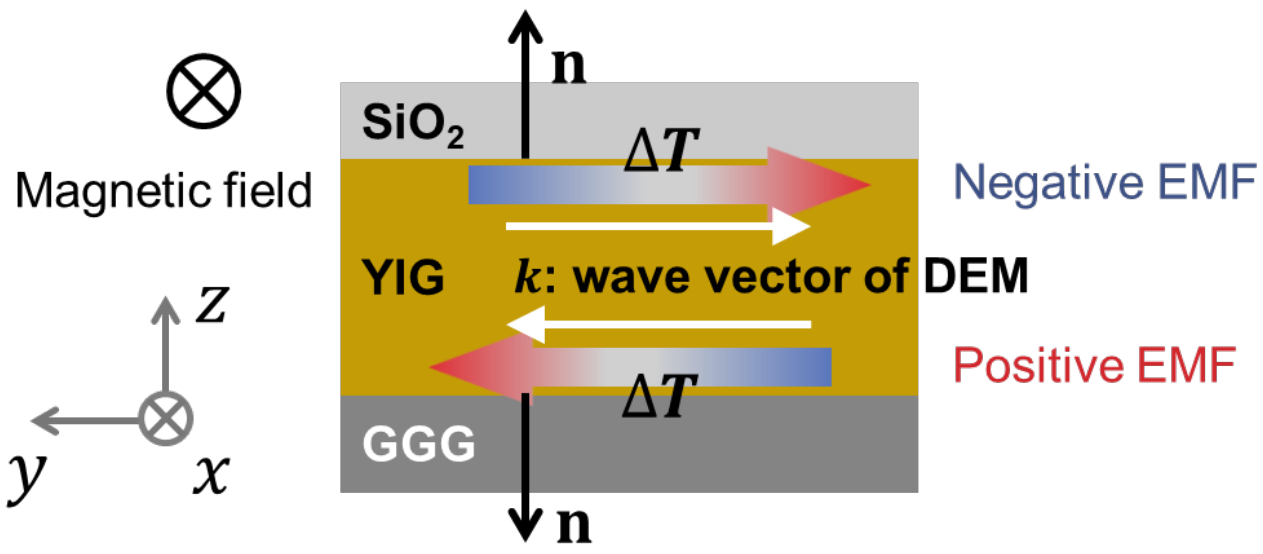

**Fig. 4.** A cross-sectional illustration of the thermal gradient generation by the DEM under the application of the external magnetic field in 0° direction. Direction of the **k** wave vector of the DEM is locked to the direction of the cross product of the YIG magnetization **M** and surface normal vector **n**.

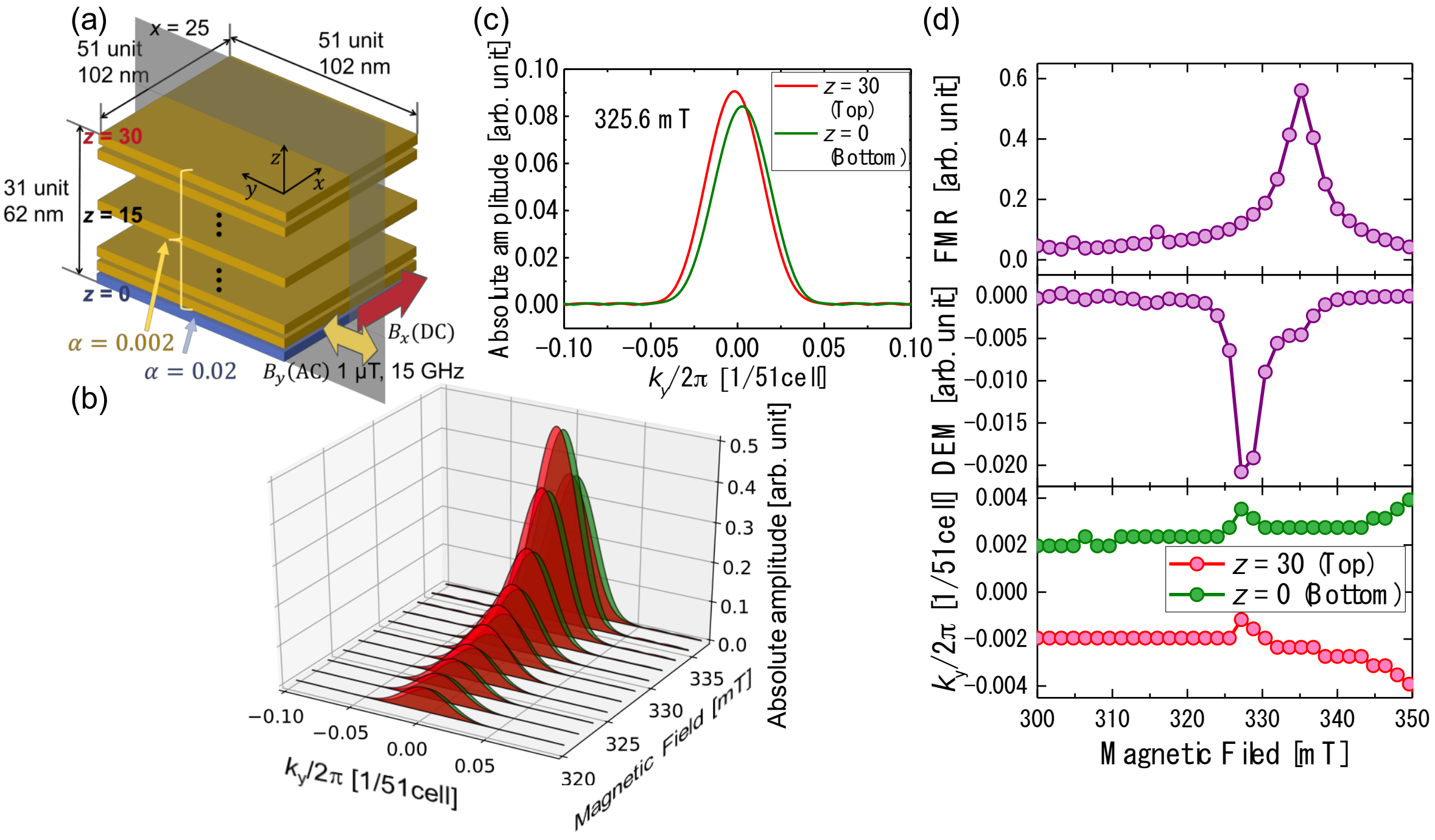

**Fig. 5.** (a) A schematic illustration of the structure used in micromagnetic simulation. The magnetic film had 31 layers in the $z$ direction, and each layer consisted of 51 × 51 unit cells. Gilbert damping parameter was set to be 0.02 in the bottom layer, and 0.002 in the other layers. The directions of the DC and AC magnetic fields are indicated by the arrows. (b) The waterfall plot of the absolute amplitude of $m_z$ component in the $x$ = 25 slice with respect to $k_y/2\pi$ and the DC magnetic field. The red and green lines indicate the top and bottom layer, respectively. (c) The absolute amplitude of $m_z$ component in the $x$ = 25 slice in the top and bottom layer at the DC magnetic field 325.6 mT. The wave form of the top (bottom) layer is biased to the -$k_y/2\pi$ (+$k_y/2\pi$) direction, indicating the propagation direction of the DEM. The amplitude of the main lobe in the top layer is higher than that in the bottom layer. (d) The upper box: the FMR intensity represented by the absolute amplitude of $m_z$ in the $z$ = 15 layer). The middle box: difference in the absolute amplitude of $m_z$ between the bottom and top layers, indicating the DEM imbalance and heat transport amplitude. The lower box: the wave numbers corresponding to the maximum of the absolute amplitude of $m_z$ in top (red) and bottom (green) layers.

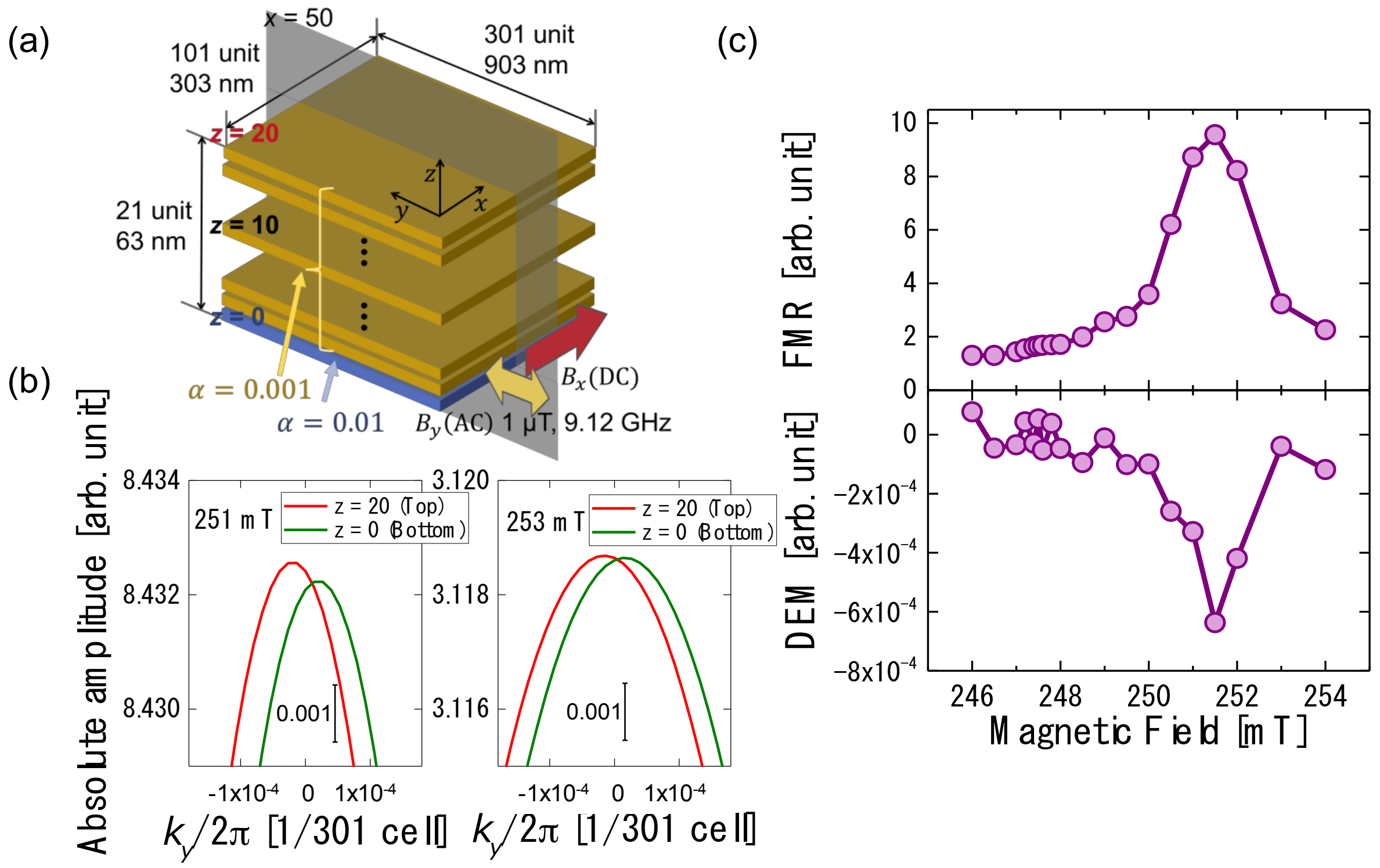

**Fig. 6.** (a) A schematic illustration of the structure used in micromagnetic simulation. The magnetic film had 21 layers in the $z$ direction, and each layer consisted of 101 × 301 unit cells. Gilbert damping was set to be 0.01 in the bottom layer, and 0.001 in the others. (b) The absolute amplitude of $m_z$ component of the magnetization in the $x$ = 50 slice in the top (red) and bottom (green) layers. The DC magnetic field was set to 251 (left box) and 253 mT (right box). (c) The upper box: The FMR intensity (the absolute amplitude of the $z$ = 15 layer). The lower box: the DEM imbalance between bottom and top layers calculated from the micromagnetic simulation. Experimental measurements indicated the similar dominance of the top DEM from the observed heat transport.